\documentclass[pre,twocolumn,showpacs,amsmath,amssymb]{revtex4}
                    
\usepackage{graphicx}
\usepackage{bm}

\begin{document}

\preprint{Working Manuscript}

\title{Coarse-grained computations of demixing in dense
gas-fluidized beds}
\author{Sung Joon Moon, S. Sundaresan, and I. G.
Kevrekidis\footnote{Electronic address: yannis@princeton.edu}}
\affiliation{Department of Chemical Engineering \&
Program in Applied and Computational Mathematics\\
Princeton University, Princeton, NJ 08544}

\date{\today}

\begin{abstract}
We use an ``equation-free'', coarse-grained computational approach
to accelerate molecular dynamics-based computations of demixing
(segregation) of dissimilar particles subject to an upward gas
flow (gas-fluidized beds).
We explore the coarse-grained dynamics of these phenomena
in gently fluidized beds of solid mixtures of
different densities, typically a slow process
for which reasonable continuum models are currently unavailable.
\end{abstract}


\maketitle


Particulate flows, even for experimental systems of small size
($\sim$ 10 cm), consist of a very large number of discrete
dissipative particles. 
Molecular dynamics (MD) simulations
often serve as a quantitative modeling tool for such flows;
however, such simulations for realistically large temporal
and/or spatial scale problems are challenging even with
modern computers.
Navier-Stokes-like, macroscopic continuum models have been
developed by many authors
based on the kinetic theory of granular materials (see
Refs.~\cite{two_fluid,goldhirsh03} and references therein);
however, {\it quantitative} continuum models for realistic
particles (accounting for frictional interactions, heterogeneity
among particles, and/or other inter-particle forces, such as
van der Waals forces) in many regimes of practical
interest (e.g. dense and/or cohesive flows where enduring
contacts between particles occur) are currently unavailable.

In this paper, we consider well-known phenomena for which
the derivation of continuum models is still in flux;
mixing and demixing (segregation) can occur when dissimilar
particle mixtures of different sizes and/or densities are
subject to a strong enough upward fluid flow~\cite{rowe7278}.
A few different continuum models, more phenomenological or more
rigorous, have been proposed~\cite{gibilaro74,vanWachem01},
which often reproduce the phenomena in a qualitatively correct
manner; however, quantitative agreement is generally
elusive~\cite{vanWachem01}.
Furthermore, kinetic theory-based continuum models for binary
mixtures are much more complicated than those for uniform
particles, and numerical simulation becomes more time-consuming
(e.g. by an order of magnitude in Ref.~\cite{vanWachem01}).
Accelerating the computation using (quantitative) microscopic
models would therefore be invaluable for such problems.
The objective of this paper is to demonstrate a multi-scale
computational approach enabling accelerated integration of
MD-based microscopic simulations of dense particulate flows.

{\it Model.}
The particles are modeled as uniform-sized soft spheres (which
can have different mass), whose inter-particle contact force
$\mathbf{F}_{cont}$ is determined following a model of Cundall
and Strack~\cite{cundall79}.
The gas phase hydrodynamics is accounted for in a volume-averaged
way~\cite{tsuji93}.
This approach has been used to study size difference-driven
demixing~\cite{feng04};  here
we consider {\em density-driven} demixing.

We deliberately choose demixing occurring in narrow beds
(cross sectional area of $15d_p \times 15d_p$ with periodic
boundary conditions for both lateral directions, where $d_p$ is
the particle diameter) as a test problem, so that the exact
results can readily be computed and used to critically test
the coarse-grained computations.
These are quasi-1D flows, where the coarse-grained gas flow is
effectively 1D, while particle simulation is fully 3D. Demixing
becomes more pronounced~\cite{formisani01} in such narrow beds.
The equation of motion for each particle can be simplified to
be~\cite{moonIECR06}:
\begin{eqnarray}
\label{oneDeq}
m_p{d\mathbf{v}_p \over dt} = && m_p\mathbf{g} + \mathbf{F}_{cont} + {V_p \over \phi}\beta(\phi) \times \nonumber \\
&& \left[(\mathbf{u}_s - \mathbf{v}_p) - {1\over (1 - \phi)^2}(\mathbf{u}_s - \mathbf{U}_s) \right],
\end{eqnarray}
where $m_p$ and $\mathbf{v}_p$ are individual particle mass and
velocity, respectively; $\mathbf{g}$ is the gravitational
acceleration;
$V_p$ is the volume of each particle;
$\phi$ is the locally-averaged solid phase volume fraction;
$\beta$ is the interphase momentum transfer
coefficient~\cite{davidson85,wen66};
$\mathbf{u}_s$ is the coarse-grained solid phase velocity;
and $\mathbf{U}_s$ is the superficial gas flow velocity,
in the direction opposite to that of the gravity.
In our study, the Reynolds number based on the particle size
is generally very small ($< \sim 0.1$), and $\beta$ is
approximated to be
\begin{equation}
\label{beta}
\beta(\phi) = 18 {\mu_g \over d_p^2} \phi (1 - \phi)^{-2.65},
\end{equation}
where 
$\mu_g$ is the gas phase viscosity.
We nondimensionalize quantities by using $\rho_s$, $d_p$,
$\sqrt{gd_p}$ and $\sqrt{d_p/g}$ as characteristic density,
length, velocity, and time scales, where $\rho_s$ is the
solid phase mass density of the lighter particles.
More details of the model can be found in
Refs.~\cite{moonIECR06,moonPF06}.

\begin{figure}[t]
\begin{center}
\includegraphics[width=.8\columnwidth]{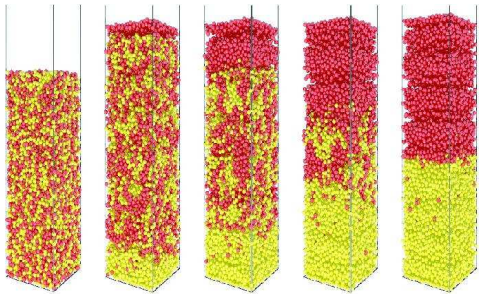}
\end{center}
\caption{\label{direct}
(Color online)
Snapshots of a gas-fluidized bed of a binary mixture of
identical size but different density particles, undergoing
spontaneous demixing, shown at times equally separated by
$\Delta t = 100.$
Light-colored (yellow) particles are twice as dense as the
dark-colored (red) ones
(coefficient of restitution (friction) = 0.9 (0.1);
$k_n = 2.0\times 10^5$; $U_s = 0.41$;
$d_p = 100~\mu$m; $\rho_s = 0.90$ g/cm$^3$).
}
\end{figure}

{\it Direct simulations.}
Demixing is typically a slow process, whose occurrence and
duration depend on the density difference and the gas flow rate.
Direct simulation with a sufficiently large $U_s$
($\equiv |\mathbf{U}_s|$),
starting from a homogeneously mixed, packed (static) state
(Fig.~\ref{direct}), illustrates that particles of different
densities gradually demix spontaneously.
When $U_s$ is well above the minimum fluidization
rate of both species (as in Fig.~\ref{direct}), the bed exhibits
1D traveling waves (TWs)~\cite{moonPF06}, and demixing occurs
superposed on the persistent oscillatory motion driven by 1D-TWs.
A typical computation of an entire demixing process in the
above ``tiny'' system ($2\times 10^7$ integration steps of 
12 500 particles shown in Fig.~\ref{direct}) takes nearly
two days, or more than a week for smaller gas flow rates,
on a single-processor PC with 1.7 GHz CPU.
Obtaining an ensemble of long simulations for statistical
averaging purposes can be extremely time-consuming.

{\it Coarse-grained description and ``observables''.}
In the literature, the degree of mixing/demixing is
often characterized by various lumped indices (or ``order
parameters'')~\cite{lacey54,rowe7278,rice86,nienow87}
such as the so-called Lacey mixing index~\cite{lacey54}
(e.g. see its use in Ref.~\cite{feng04}).
Here we seek coarse-grained variables (or ``observables'')
that could be used in continuum descriptions.

As in the two-fluid modeling approach~\cite{two_fluid},
it is natural to think of hydrodynamic variables as candidate
coarse observables.
From direct simulations, we observe that in the course of demixing
in quasi-1D beds, the process strongly depends on the local density,
which makes the 1D volume fraction profiles themselves sufficient
coarse observables;
when we suddenly randomize only the individual particle
velocities (hence the granular temperature as well)
and continue the simulation, the demixing progresses 
essentially undisturbed.
We further recognize that cumulative particle distribution
functions (CDFs, and more precisely, their inverses which
are bounded by 0 and 1) are more convenient coarse observables:
CDFs are smoother than volume fraction profiles, suffer from
less noise, and facilitate the {\it lifting} (see below) procedure.

\begin{figure}[t]
\begin{center}
\includegraphics[width=.75\columnwidth]{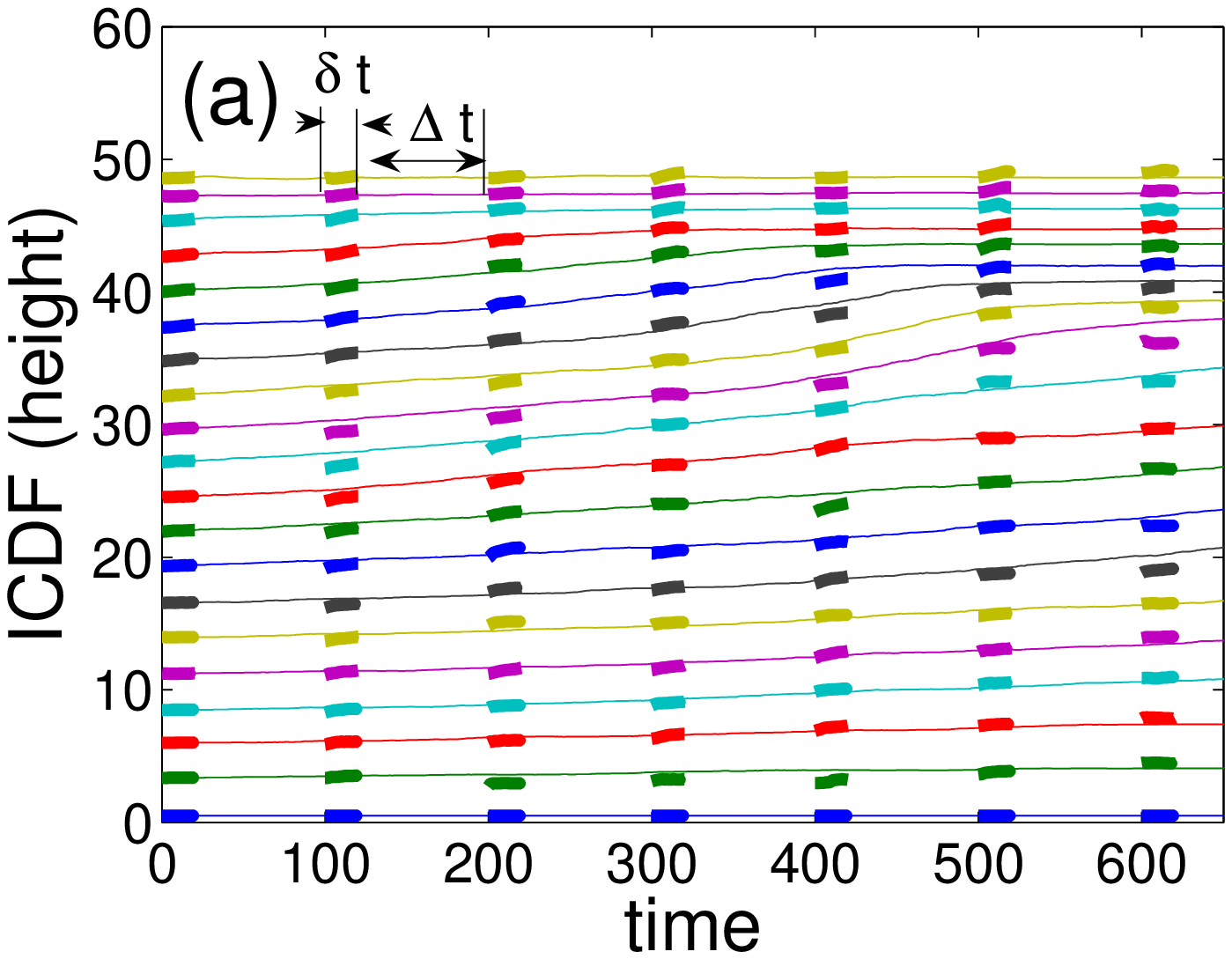}
\includegraphics[width=.75\columnwidth]{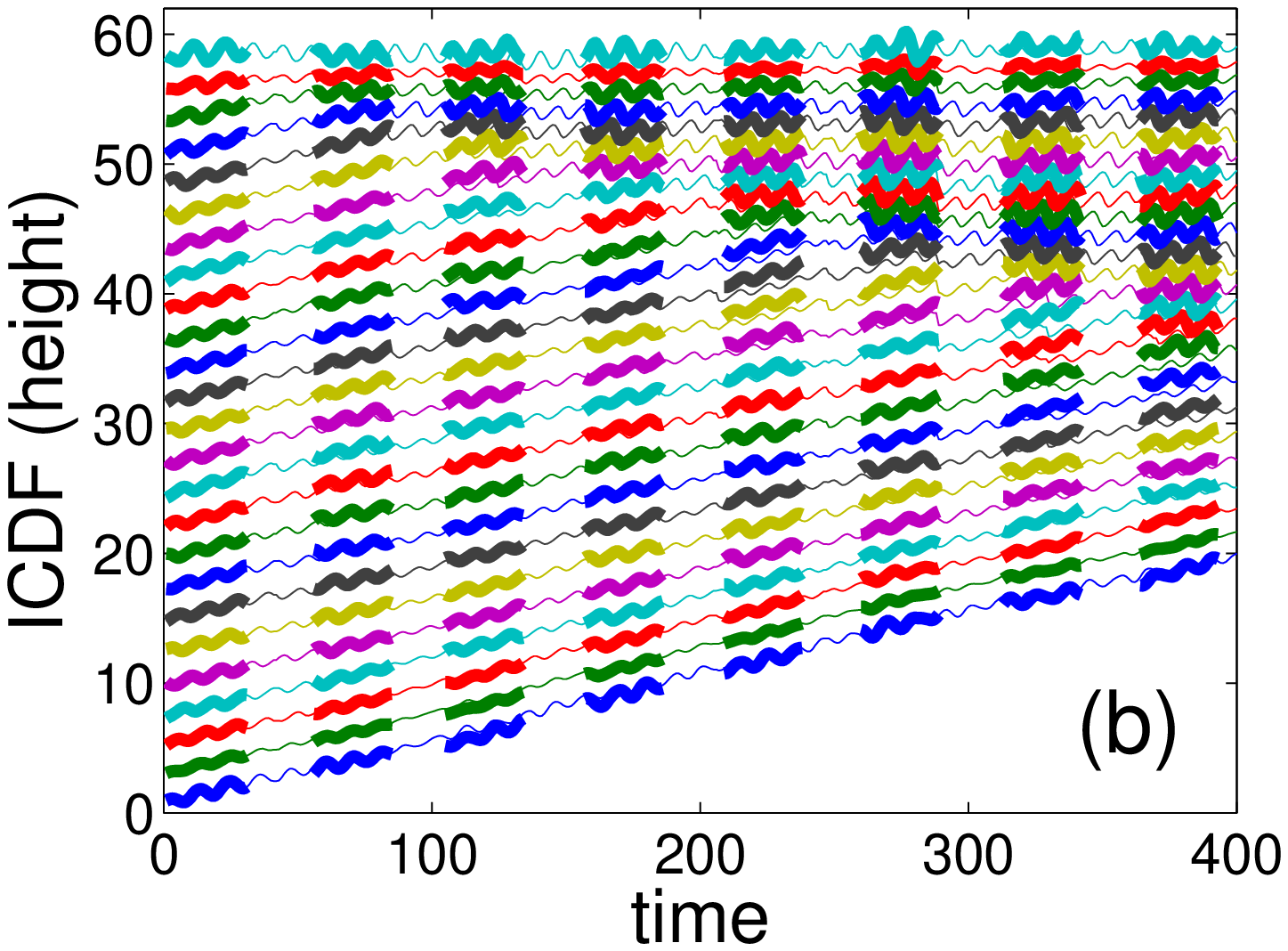}
\end{center}
\caption{\label{PIsingle}
(Color online)
Direct, full integration (thin lines) and coarse projective
integration (patches of thick lines), using 
evenly discretized CDFs as coarse observables.
(a) Slow demixing at $U_s = 0.19$, where the bed
hardly expands and no 1D-TWs form.
Projective steps of $\Delta t$ = 80 were taken through forward
Euler, after direct integration for $\delta t$ = 20
(the latter half of data were used to estimate the local slope).
(b) In the presence of 1D-TWs ($U_s = 0.41$);
$\delta t$ = 30, and $\Delta t = 2T$, where $T$ is the average
period estimated during short bursts of direct integrations;
the last two periods of the locally oscillating data were
used to estimate the coarse slope.
}
\end{figure}

\begin{figure}[t]
\begin{center}
\includegraphics[width=.75\columnwidth]{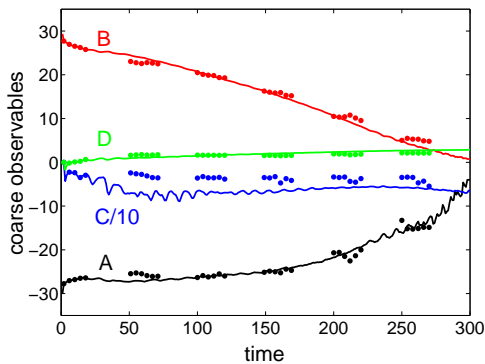}
\end{center}
\caption{\label{PIdiff}
(Color online)
Comparison between direct, full integrations (solid lines)
and coarse projective integrations (groups of dots),
using the four coarse observables in Eq.~(\ref{fitting}).
}
\end{figure}

We consider both discretized inverse CDFs (ICDFs; Fig.~\ref{PIsingle})
and their compact parametric representations (finite element or
expansion coefficients in convenient polynomial sets;
Fig.~\ref{PIdiff}) as our actual coarse observables.
For a narrow range of $U_s$'s slightly above the
minimum fluidization rate,
the time evolution of ICDFs does not exhibit any waviness
(thin lines in Fig.~\ref{PIsingle} (a)), and demixing occurs
very slowly (complete demixing is hardly achieved).
In the presence of 1D-TWs at larger $U_s$'s, ICDFs
locally oscillate at a fast time scale (Fig.~\ref{PIsingle} (b));
these oscillations
can be smoothed through ensemble-averaging.
Once the coarse observables are chosen, governing equations for
their time evolution need to be derived for further computation.
We will follow an {\it equation-free} approach, circumventing
the derivation of such equations, assuming they do exist
conceptually, but are not available explicitly.

{\it An equation-free approach.}
When the time series of coarse observables (obtained by direct
integration of the microscopic simulator) are smooth and slowly
varying, one can estimate their local time derivatives and then
project the values at a future
time (e.g. using forward Euler or more sophisticated schemes).
We recognize that {\it if} one can initialize the microscopic
simulator consistent with the future (projected) values of the
coarse observables,
one can actually accelerate the overall computation.
This simple idea underpins {\it coarse projective
integration}~\cite{gear02,gear03}.

In equation-free computations, traditional continuum
numerical techniques are directly applied to the outcome of
appropriately initialized short bursts of microscopic
simulation, and the macroscopic equations are
``integrated'' or ``solved'' without ever being written
down~\cite{pnas00,review03,smallmanifesto}.
The essential steps are:
(i) Identify coarse observables (which are discretized
ICDFs or their parameterization coefficients in our study).
For convenience, we denote the microscopic
description (here the individual particle positions)
by $\mathbf{x}$, and the macroscopic description
(here the ICDFs) by $\mathbf{X}$.
(ii) Choose an appropriate {\it lifting} operator $\mu_L$,
which maps $\mathbf{X}$ (ICDFs) to one (or more, for the
purposes of variance reduction and ensemble-averaging,)
consistent description(s) $\mathbf{x}$ (here, particle positions).
Figuring out an efficient lifting operator is essential.
(iii) Starting from lifted initial condition(s)
$\mathbf{x}(t_0) = \mu_L(\mathbf{X}(t_0))$,
run the detailed simulator for some time horizon ($T_h > 0$)
to obtain $\mathbf{x}(t_0+T_h)$.
(iv) Use an appropriate {\it restriction} operator $\mathcal{M}_R$
which maps the microscopic state(s) to the macroscopic description
$\mathbf{X}(t_0+T_h) = \mathcal{M}_R(\mathbf{x}(t_0+T_h))$, resulting
in time series of the coarse observables (ICDFs), 
or a coarse time-stepper $\Phi_{T_h}$
for them: $\mathbf{X}(t_0+T_h) \equiv \Phi_{T_h}(\mathbf{X}(t_0))$.
(v) Apply desired numerical techniques (forward Euler, in our study)
to the coarsely observed results in step (iv), and repeat.

{\it Lifting.}
Given an ICDF as the coarse observable, we
need to construct particle configurations consistent with it.
Arranging particles (i.e. sphere packing) in
{\it three-dimensional} space with an {\it arbitrarily} prescribed
ICDF (or local volume fraction, especially dense) profile is
nontrivial and generally time-consuming~\cite{torquato02}, as
excessive particle-particle overlap has to be avoided;
it would be less difficult for dilute particulate flows.
In our study, the total height of a bed
remains virtually the same, even in the presence
of 1D-TWs (see the second and later frames in Fig.~\ref{direct}).
Therefore, we do not reassign particle locations from scratch in each
lifting step. Instead, we utilize particle locations
which are already obtained from an earlier step, and switch
{\it only} particle indices (or ``colors'', whether ``red''
or ``yellow'') until the prescribed ICDF becomes satisfied.
When a system of polydisperse particles is considered, this
scheme should be modified; this would be a subject of future
research.

Our lifting operator for single realization computations
requires additional consideration when 1D-TWs are present:
particle locations obtained from earlier simulation at
the {\it same} ``phase angle'' during the wave propagation
have to be used.
In ensemble-averaged computations, the oscillatory motion
becomes smoothed, requiring no special attention.

\begin{figure}[t]
\begin{center}
\includegraphics[width=.7\columnwidth]{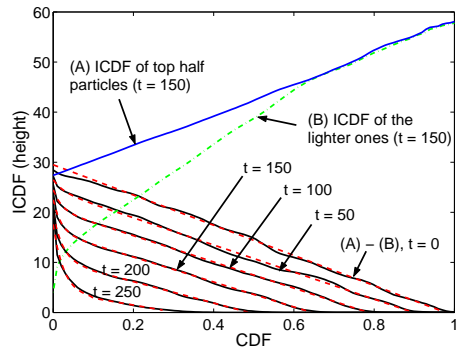}
\end{center}
\caption{\label{CDFdiff}
(Color online)
Evolving inverse CDFs (ICDFs) for the same case as in
Fig.~\ref{direct}, averaged over 10 realizations:
(A) ICDF of particles physically located in the upper half
of the bed, and (B) ICDF of the lighter particles, which are
shown at $t = 150$.
Solid lines with negative slopes are snapshots of the difference
between (A) and (B) at different times; compare with
the functional fit shown as dashed lines; see Eq.~(\ref{fitting}).
}
\end{figure}

{\it Coarse projective integration.}
We choose discretized ICDFs of the lighter particles as the
coarse observables, and accelerate the demixing computation
using coarse projective forward Euler scheme~\cite{gear02,gear03}.
For smaller $U_s$, where the bed hardly expands and ICDFs do
not oscillate, the projection step size is determined by only
the temporal smoothness (accuracy of the local linearization)
of ICDF evolution.
The demixing occurs very slowly in this case (Fig.~\ref{PIsingle} (a)),
and these computations can achieve high computational speedup.
%
Excessively large projection steps can cause inaccuracies,
similar to large time steps in normal integration.
In the presence of 1D-TWs, the projection step size is chosen to be
an integral multiple of the local oscillation period
(Fig.~\ref{PIsingle} (b)).

Projectively integrated values (thick lines in Fig.~\ref{PIsingle})
follow the trajectories of direct, full integrations (thin
lines in Fig.~\ref{PIsingle}) well, confirming that these
coarse observables are good continuum variables.
Ensemble-averaging of ICDFs over different realizations (of
different phase angles during wave propagation) smoothens
local oscillations arising from 1D-TWs.
Projective integration of ensemble-averaged ICDFs, both in
the presence and absence of 1D-TWs, can be applied in the same
way.

{\it A more compact description.}
The difference between the ICDF of particles
located in the {\em upper half} of the bed (irrespective of their
densities) and that of the lighter ones can serve as a useful coarse
observable;
ICDFs of the lighter particles are bounded by those of the
(total) top half particles, and their difference is positive
definite (Fig.~\ref{CDFdiff}). 
Furthermore, once nearly-full local demixing occurs,
the ICDF difference there becomes virtually zero.
The difference of the two ICDFs can be fit by the following simple
functional form (dashed lines in Fig.~\ref{CDFdiff}):
\begin{equation}
 \label{fitting}
y = {\rm max}[Ax + B + C\exp(Dx),0],
\end{equation}
where $A(t), B(t), C(t)$, and $D(t)$ are our new coarse
observables, determined on the fly by functional fitting,
and $x$ and $y$ represent the abscissa and the ordinate
in Fig.~\ref{CDFdiff} respectively.
Other basis functions, such as high-order polynomials also
can fit the functional form reasonably well, but they require
many more terms and the time evolution of their expansion
coefficients is not generally slow.
Lifting for these new coarse observables involves
a minor intermediate step: mapping between these four
variables and an ICDF discretization through the functional
form in Eq.~(\ref{fitting}).

We use these four coarse observables to perform ensemble-averaged
coarse projective integration over a number of realizations
(Fig.~\ref{PIdiff}).
These observables vary slowly and smoothly in time
(occasional oscillations disappear for larger 
ensembles);
in a sense, this is a pseudospectral solution of the
unknown governing equations for ICDF evolution.

{\it Conclusions.}
We have used an ``equation-free'' coarse-grained approach to
accelerate (by a factor of two to ten; the lifting step in our
study involves minimal computational effort) computations of
dense particulate flows and, in particular, of demixing
occurring in gas-fluidized beds of dissimilar particles.
This approach holds promise for the prediction of coarse-grained
behavior at practically relevant spatial and temporal scales.

We deliberately considered a quasi-1D illustrative problem
in our study, in order to demonstrate the viability of the approach.
As a consequence of the problem considered in this study,
the coarse observables were {\it one-dimensional} discretized ICDFs.
For systems involving higher dimensional flows, candidates for
coarse observables may include marginal and conditional
ICDFs~\cite{zou06}.
More work for such systems has to be done to identify proper
coarse observables and an efficient lifting operator, 
vital components of this approach.
Ensemble averaging reduces fluctuations among the realizations,
giving better quantitative representations;
the computation of each realization 
readily parallelizes across computational
nodes.

More sophisticated equation-free algorithms (e.g. coarse fixed point
algorithms~\cite{review03}) can be used to find stable as
well as unstable steady states;
quantify their stability; and perform numerical
bifurcation analysis. Exploiting such tools to 
investigate the coarse-grained dynamics of mixing
and demixing (and other particulate flow problems)
is the subject of current research.

This research was partially supported by
DOE, DARPA, and ACS-PRF.

\end{document}